# A New Round Robin Based Scheduling Algorithm for Operating Systems: Dynamic Quantum Using the Mean Average

**Abbas Noon[1], Ali Kalakech[2], Seifedine Kadry[1]**

[1] Faculty of Computer Science, Arts Sciences and Technology University
Lebanon

[2] Faculty of Business, Lebanese University
Lebanon

**Abstract**

Round Robin, considered as the most widely adopted CPU scheduling algorithm, undergoes severe problems directly related to quantum size. If time quantum chosen is too large, the response time of the processes is considered too high. On the other hand, if this quantum is too small, it increases the overhead of the CPU.

In this paper, we propose a new algorithm, called AN, based on a new approach called dynamic-time-quantum; the idea of this approach is to make the operating systems adjusts the time quantum according to the burst time of the set of waiting processes in the ready queue.

Based on the simulations and experiments, we show that the new proposed algorithm solves the fixed time quantum problem and increases the performance of Round Robin.

## 1. Introduction

Modern Operating Systems are moving towards multitasking environments which mainly depends on the CPU scheduling algorithm since the CPU is the most effective or essential part of the computer. Round Robin is considered the most widely used scheduling algorithm in CPU scheduling [8, 9], also used for flow passing scheduling through a network device [1].

CPU Scheduling is an essential operating system task, which is the process of allocating the CPU to a specific process for a time slice. Scheduling requires careful attention to ensure fairness and avoid process starvation in the CPU. This allocation is carried out by software known as scheduler and dispatcher [8, 9].

Operating systems may feature up to 3 distinct types of a long-term scheduler (also known as an admission scheduler or high-level scheduler), a mid-term or medium-term scheduler and a short-term scheduler (fig1).

The dispatcher is the module that gives control of the CPU to the process selected by the short-term scheduler [8].

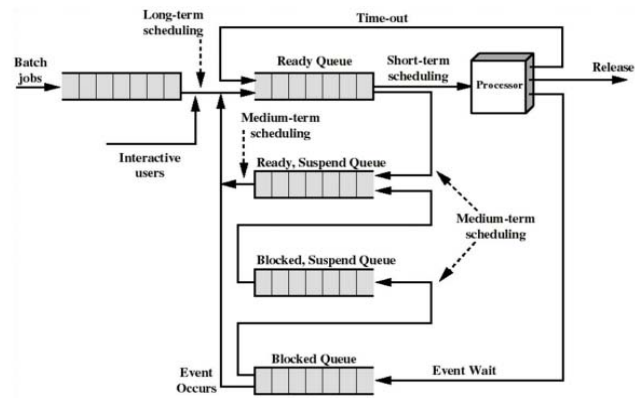

Figure 1: Queuing diagram for scheduling

There are many different scheduling algorithms which varies in efficiency according to the holding environments, which means what we consider a good scheduling algorithm in some cases which is not so in others, and vice versa. The Criteria for a good scheduling algorithm depends, among others, on the following measures [8]:

- Fairness: all processes get fair share of the CPU,
- Efficiency: keep CPU busy 100% of time,
- Response time: minimize response time,
- Turnaround: minimize the time batch users must wait for output,
- Throughput: maximize number of jobs per hour.

Moreover, we should distinguish between the two schemes of scheduling: preemptive and non preemptive algorithms. Preemptive algorithms are those where the burst time of a process being in execution is preempted when a higher priority process arrives. Non preemptive algorithms are used where the process runs to complete its burst time even a higher priority process arrives during its execution time.

First-Come-First-Served (FCFS)[8, 9] is the simplest scheduling algorithm, it simply queues processes in the order that they arrive in the ready queue. Processes are dispatched according to their arrival time on the ready queue. Being a non preemptive discipline, once a process has a CPU, it runs to completion. The FCFS scheduling is fair in the formal sense or human sense of fairness but it is unfair in the sense that long jobs make short jobs wait and unimportant jobs make important jobs wait [8, 9].

Shortest Job First (SJF) [8, 9] is the strategy of arranging processes with the least estimated processing time remaining to be next in the queue. It works under the two schemes (preemptive and non-preemptive). It's provably optimal since it minimizes the average turnaround time and the average waiting time. The main problem with this discipline is the necessity of the previous knowledge about the time required for a process to complete. Also, it undergoes a starvation issue especially in a busy system with many small processes being run [8, 9].

Round Robin (RR) [8, 9]which is the main concern of this research is one of the oldest, simplest and fairest and most widely used scheduling algorithms, designed especially for time-sharing systems. It's designed to give a better responsive but the worst turnaround and waiting time due to the fixed time quantum concept. The scheduler assigns a fixed time unit (quantum) per process usually 10-100 milliseconds, and cycles through them. RR is similar to FCFS except that preemption is added to switch between processes [2, 3, and 8].

In this paper, we propose a new algorithm to solve the constant time quantum problem. The algorithm is based on dynamic time quantum approach where the system adjusts the time quantum according to the burst time of processes founded in the ready queue. The second section states some of previous works done in this field. Section III describes the proposed method in details. Section IV discusses the simulation done in this method, before concluding this paper in the last section.

## 2. Previous works

Round Robin becomes one of the most widely used scheduling disciplines despite of its severe problem which rose due to the concept of a fixed pre-determined time quantum [2, 3, 4, 5, 6, and 7]. Since RR is used in almost every operating system (windows, BSD, UNIX and Unix-based etc…), many researchers have tried to fill this gap, but still much less than needs.

Matarneh [2] founded that an optimal time quantum could be calculated by the median of burst times for the set of processes in ready queue, unless if this median is less than 25ms. In such case, the quantum value must be modified to 25ms to avoid the overhead of context switch time [2]. Other works [7], have also used the median approach, and have obtained good results.

Helmy et al. [3] propose a new weighting technique for Round-Robin CPU scheduling algorithm, as an attempt to combine the low scheduling overhead of round robin algorithms and favor short jobs. Higher process weights means relatively higher time quantum; shorter jobs will be given more time, so that they will be removed earlier from the ready queue [3]. Other works have used mathematical approaches, giving new procedures using mathematical theorems [4].

Mohanty and others also developed other algorithms in order to improve the scheduling algorithms performance [5], [6] and [7]. One of them is constructed as a combination of priority algorithm and RR [5] while the other algorithm is much similar to a combination between SJF and RR [6].

## 3. AN Algorithm

In this paper, we present a solution to the time quantum problem by making the operating system adjusts the time quantum according to the burst time of the existed set of processes in the ready queue.

3.1 Methodology

When operating system is installed for the first time, it begins with time quantum equals to the burst time of first dispatched process, which is subject to change after the end of the first time quantum. So, we assume that the system will immediately take advantage of this method.
The determined time quantum represents real and optimal value because it based on real burst time unlike the other methods, which depend on fixed time quantum value. Repeatedly, when a new process is loaded into the ready queue in order to be executed, the operating system calculates the average of sum of the burst times of processes found in the ready queue including the new arrival process.
This method needs two registers to be identified:
- SR: Register to store the sum of the remaining burst times in the ready queue.

- AR: Register to store the average of the burst times by dividing the value found in the SR by the count of processes found in the ready queue.

When a process in execution finishes its time slice or its burst time, the ready queue and the registers will be updated to store the new data values.
- If this process finishes its burst time, then it will be removed from the ready queue. Otherwise, it will move to the end of the ready queue.
- SR will be updated by subtracting the time consumed by this process.
- AR will be updated according to the new data.

When a new process arrives to the ready queue, it will be treated according to the rules above in addition to updating the ready queue and the registers.

### 3.2 Pseudo Code and Flow Chart

The algorithm described in the previous section can be formally described by pseudo code and flow chart like follows:

```
New process P arrives
 P Enters ready queue
Update SR and AR
Process p is loaded from ready queue
into the CPU to be executed
  IF (Ready Queue is Empty)
    TQ  BT (p)
    Update SR and AR
  End if
  IF (Ready Queue is not empty)
    TQAVG (Sum BT of processes in
ready queue)
    Update SR and AR
  End if
CPU executes P by TQ time
  IF (P is terminated)
      Update SR and AR
  End if
  IF (P is not terminated)
     Return p to the ready queue with
its updated burst time
     Update SR and AR
  End if
```

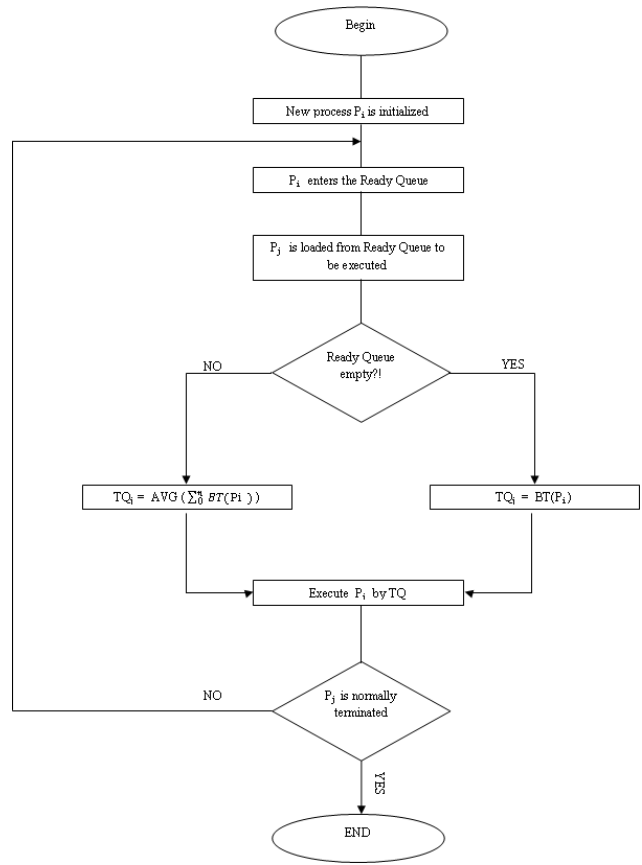

## 4. Simulations

In order to validate our algorithm (AN) over the existing Round Robin, we have built our simulator using MATLAB, since it presents the user data and solutions after fetching in a graphical representation which is not found in most other languages.

Using MATLAB 2010a, we built a simulator for AN algorithm that acquires a triplet (N, AT, BT) where:
- N: the number of processes
- AT: an array of arrival times of all processes
- BT: an array of burst times of all processes

The simulator calculates the average waiting time and the average turnaround time of the whole system consisting of N processes according to the AN algorithm.

We have also built a simulator for Round Robin algorithm that acquires a quadrant (Q, N, AT, BT) where:
- Q: The time quantum (assigned by the user)
- N: the number of processes
- AT: an array of arrival times of all processes
- BT: n array of burst times of all processes

Then the simulator calculates the average waiting time and the average turnaround time of the whole system consisting of N processes according to the Round Robin algorithm.

Finally, we have developed a simple function to compare among the two algorithms presenting graphical result, showing the efficiency of our algorithm over Round Robin. The function loads data from a text file consisting of 50 samples. Each sample is a 4 processes system (N=4). Arrival times and burst times were randomly chosen varying from 10 To 100 milliseconds. Note that we choose N = 4 since whatever N is, we will have the same result as will shown in the result below (figures 2 and 3).

We have chosen a fixed time quantum Q=10 ms in Round Robin it gives the results in fig2 and fig3. In these figures, the x-axis represents the different samples we have targeted, while the y-axis represents the TAT (average of turnaround times) in fig 2, and the WT (average of waiting times) in fig3. In the graphs below a higher vertex means a larger average turnaround time (fig2) and waiting time (fig3). As mentioned before a better algorithm is to minimize turnaround and waiting time, thus the better algorithm has the lowest vertex.

These figures clearly show that for all the tested cases, we obtain better results (lower TAT and WT) when using the AN algorithm.

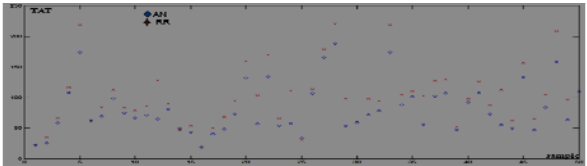
Figure 2: Average Turnaround time for time quantum = 10 ms

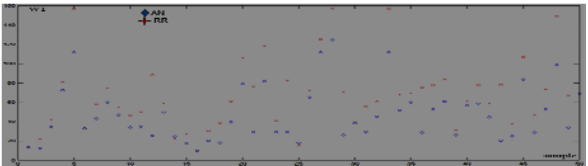
Figure 3: Average Waiting time for time quantum = 10 ms

The same process was done on TQ=15, 20, 25 and 30 ms to cover as much as possible fixed time quantum possibilities, and we always obtain the same results.

## 4. Results and Observations

As a result of the simulation and hand solved examples we've reached to a conclusion that AN algorithm could improve the efficiency of Round Robin by changing the idea of fixed time quantum to dynamic calculated automatically without the interfere of user.

### 4.1 Numerical Examples

To evaluate our proposed method and for simplicity seek we will take a group of four processes in four different cases with random burst, in fact the number of processes does not change the result because the algorithm works effectively even if it used with a very large number of processes. For each case, we will compare the result of our developed method with the traditional approach (fixed quantum = 20ms) and with the method proposed in [2]. We should mention here, the numerical values of the 4 different cases are taken from [2].

**Case 1**: Assume four processes arrived at time = 0, with burst time (P1 = 20, P2 = 40, P3 = 60, P4 = 80):

|  | Fixed Quantum=20ms | Dynamic method [2] | AN |
|---|---|---|---|
| Turn-around time | 120 | 112.5 | 100 |
| Waiting time | 70 | 77.5 | 50 |
| Context switch | 9 | 6 | 5 |

**Case 2**: Assume four processes arrived at time = 0, with burst time (P1 = 10, P2 = 14, P3 = 70, P4 = 120):

|  | Fixed Quantum=20ms | Dynamic method [2] | AN |
|---|---|---|---|
| Turn-around time | 100.5 | 96 | 85.5 |
| Waiting time | 47 | 42.5 | 32 |
| Context switch | 11 | 6 | 5 |

**Case 3**: Assume four processes arrived at different time, respectively 0, 4, 8, and 16, with burst time (P1 = 18, P2 = 70, P3 = 74, P4 = 80):

|  | Fixed Quantum=20ms | Dynamic method [2] | AN |
|---|---|---|---|
| Turn-around time | 106 | 98.5 | 81 |
| Waiting time | 60 | 58.5 | 35 |
| Context switch | 10 | 4 | 5 |

**Case 4**: Assume four processes arrived at different time, respectively 0, 6, 13, and 21, with burst time (P1 = 10, P2 = 14, P3 = 70, P4 = 120):

|  | Fixed Quantum 20ms | Dynamic method [2] | AN |
|---|---|---|---|
| Turn-around time | 90.5 | 46 | 75.5 |
| Waiting time | 37 | 30.5 | 22 |
| Context switch | 11 | 4 | 4 |

From the above comparisons, it is clear that the dynamic time quantum approach based on the average of processes bursts time is more effective than the fixed time quantum approach and the proposed method in [2] in round robin algorithm, where the dynamic time quantum significantly reduces the context switch, turnaround time and the waiting time. In addition, the complexity calculation of the mean of the processes is very small.

### 4.2 Improvements in waiting times and turnaround times

At the end of each run we calculated the percentage of improvement of AN algorithm over Round Robin by implementing a simple rule.
I = (Vertex [AN] – Vertex [RR])/number of samples
We obtained the following results (table 1):

**Table 1: Improvement percentage of AN**

| TQ | % I(wt[TQ]) | % I(tat[TQ]) |
|---|---|---|
| 10 ms | 20.1162 | 20.1162 |
| 15 ms | 16.1163 | 16.1162 |
| 20 ms | 13.8562 | 13.8562 |
| 25 ms | 12.6113 | 12.6112 |
| 30 ms | 10.4413 | 10.4412 |

### 4.3 Success in Statistics

In addition to the improvement measure (%I), we added another measure of success over failure which is calculated by percentage of success samples over the failed ones. A succeed sample is sample where vertex of AN algorithm is less than vertex of RR.
S= ((number of succeed samples) / (total number of samples)) we obtained the following results (table 2).

**Table 2: Success over failure percentage of AN**

| TQ | %S(tat[TQ]) | %S(wt[TQ]) |
|---|---|---|
| 10 ms | 96% | 96% |
| 15 ms | 92% | 90% |
| 20 ms | 90% | 88% |
| 25 ms | 88% | 88% |
| 30 ms | 86% | 84% |

### 4.4 Improvement in Context Switches

As a result of our observations, 50% of the processes will be terminated through the first round and as time quantum is calculated repeatedly for each round, then 50% of the remaining processes will be terminated during the second round, with the same manner for the third round, fourth round etc…i.e., the maximum number of rounds will be less than or equal to 6 whatever the number of processes or their burst time (fig4). [2]

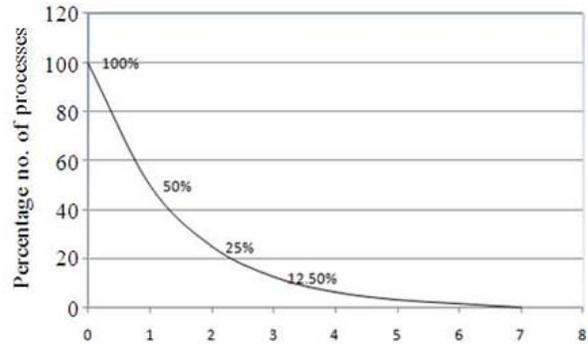
Figure 4: The rate of decrease in the number of processes in each round

The significant decrease of the number of processes will inevitably lead to significant reduction in the number of context switches, which may pose high overhead on the operating system in many cases. The number of context switches can be represented mathematically as follows:

$$Q_T = \left(\sum_1^r K_r\right) - 1$$

Where:
QT = the total number of context switch
r = the total number of rounds, r = 1, 2…6
$k_r$ = the total number of processes in each round

In other variants of round robin scheduling algorithm, the context switch occurs even if there is only a single process in the ready queue, where the operating system assigns to the process a specific time quantum Q[4]. When time quantum expires, the process is interrupted and again assigned the same time quantum Q, regardless whether the process is alone in the ready queue or not [2, 3], which means that there will be additional unnecessary context switches, while this problem does not occur at all in our new proposed algorithm; because in this case, the time quantum will equal to the remaining burst time of the process.

## 5. Conclusion

Time quantum is the bottleneck facing round robin algorithm and was more frequently asked question: What is the optimal time quantum to be used in round robin algorithm?
In light of the effectiveness and the efficiency of the RR algorithm, this paper provides an answer to this question by using dynamic time quantum instead of fixed time quantum, where the operating system itself finds the optimal time quantum without user intervention.
In this paper, we have discussed the AN algorithm that could be a simple step for a huge aim in obtaining an optimal scheduling algorithm. It will need much more efforts and researches to score a goal.

From the simulation study, we get an important conclusion; that the performance of AN algorithm is higher than that of RR in any system. The use of dynamic scheduling algorithm increased the performance and stability of the operating system and supports building of a self-adaptation operating system, which means that the system is who will adapt itself to the requirements of the user and not vice versa.